\documentclass[journal]{IEEEtran}

\usepackage{ifpdf}
\usepackage{graphicx}
\usepackage[space]{grffile}
\usepackage{latexsym}
\usepackage{textcomp}
\usepackage{longtable}
\usepackage{tabulary}
\usepackage{booktabs,array,multirow}
\usepackage{amsfonts,amsmath,amssymb}
\providecommand\citet{\cite}
\providecommand\citep{\cite}

\usepackage{url}
\usepackage{hyperref}
\hypersetup{colorlinks=false,pdfborder={0 0 0}}
\usepackage{etoolbox}
\newif\iflatexml\latexmlfalse

\AtBeginDocument{\DeclareGraphicsExtensions{.pdf,.PDF,.eps,.EPS,.png,.PNG,.tif,.TIF,.jpg,.JPG,.jpeg,.JPEG}}

\usepackage[utf8]{inputenc}
\usepackage[ngerman,english]{babel}
\usepackage{cite}
\begin{document}

\title{Using Deep Learning with Large Aggregated Datasets for COVID-19 Classification from Cough}

%
%
%

\author{Esin Darici Haritaoglu,
Nicholas Rasmussen,
Daniel C. H. Tan,
Jennifer Ranjani J.,
Jaclyn Xiao,
Gunvant Chaudhari,
Akanksha Rajput,
Praveen Govindan,
Christian Canham, 
Wei Chen,
Minami Yamaura,
Laura Gomezjurado,
Amil Khanzada,
Aaron Broukhim,
Mert Pilanci
\thanks{Esin Darici Haritaoglu is with Virufy}

\thanks{Nicholas Rasmussen is with the Department of Computer Science, University of South Dakota}
\thanks{Daniel C. H. Tan is with the Department of Computer Science, Stanford University}
\thanks{Jennifer Ranjani J. is with Virufy}
\thanks{Jaclyn Xiao is with the Department of Biomedical Engineering, Duke University}
\thanks{Gunvant Chaudhari is with the School of Medicine, University of California San Francisco}
\thanks{Akanksha Rajput is with the Data Science Institute, Columbia
University}
\thanks{Praveen Govindan is with Virufy}
\thanks{Christian Canham is with Virufy}
\thanks{Wei Chen is with University of California, Riverside}
\thanks{Minami Yamaura is with the Electrical Engineering and Computer Sciences Department, University of California Berkeley}
\thanks{Laura Gomezjurado is with Virufy}
\thanks{Amil Khanzada is with the Electrical Engineering and Computer Sciences Department and Haas MBA, University of California, Berkeley}
\thanks{Aaron Bourkhim is with the Department of Computer Science and Engineering, University of California, San Diego}
\thanks{Mert Pilanci is with the Department of Electrical Engineering, Stanford University}
\thanks{Esin Darici Haritaoglu and Nicholas Rasmussen have contributed equally to this work}
\thanks{Mert Pilanci, Ph. D. is supported by the National Science Foundation (NSF), and the U.S. Army Research Office}
}


\markboth{}%
{Esin Darici {et~al.}: Using Deep Learning with Large Aggregated Datasets}

\maketitle


\selectlanguage{english}
\begin{abstract}
The Covid-19 pandemic has been one of the most devastating events in recent history, claiming the lives of more than 5 million people worldwide. Even with the worldwide distribution of vaccines, there is an apparent need for affordable, reliable, and accessible screening techniques to serve parts of the World that do not have access to Western medicine. Artificial Intelligence can provide a solution utilizing cough sounds as a primary screening mode for COVID-19 diagnosis. This paper presents multiple models that have achieved relatively respectable performance on the largest evaluation dataset currently presented in academic literature. Through investigation of a self-supervised learning model (Area under the ROC curve, AUC = 0.807) and a convolutional nerual network (CNN) model (AUC = 0.802), we observe the possibility of model bias with limited datasets. Moreover, we observe that performance increases with training data size, showing the need for the worldwide collection of data to help combat the Covid-19 pandemic with non-traditional means.
\end{abstract}%


\IEEEpeerreviewmaketitle

\section{Introduction}
\label{intro} 
To date, COVID-19 has claimed the lives of over 800,000 individuals in the United States and 5 million Worldwide \cite{dashboard}. The highly infectious nature of COVID-19 has filled up hospital beds in record numbers, surpassing hospital capacity and causing immense strain on healthcare systems Worldwide \cite{wadhera2021}. While global vaccination efforts are underway, distribution efforts have been impeded in low and middle-income countries. Additionally, the emergence of new viral variants like the Omicron has decreased the effectiveness of current vaccines in the prevention of COVID-19 spread \cite{Mahase2021}.

One of the main contributors to the spread of COVID-19 is that a proportion of infected people show mild or moderate symptoms while infectious \cite{Ma2021}. There is a high demand for frequent, rapid, and affordable testing. To this date, reverse transcription-polymerase chain reaction (RT-PCR) tests in nasopharyngeal swabs are the gold standard for detecting COVID-19 in clinical practice due to their high sensitivity and specificity \cite{kruger2021}. However, despite the reliability of RT-PCR tests, some issues have arisen during mass application. One such example is that these tests require costly reagents and tools that have prevented equal access globally. Furthermore, administering and processing the test poses a risk of possible infection, and the test results take hours to days to return. An alternative is an antigen detection test for COVID-19 such as the COVID-19 Ag Respi-Strip, a rapid immunochromatographic test for detecting SARS-CoV-2 antigen. This test has the ability to deliver results in as little as 15 minutes, a factor that is helpful in hospital diagnosis for the safety of its patients. However, the antigen test has shown low sensitivity, with studies showing that its sensitivity was as low as 30.2\% \cite{Scohy2020}, and there is often limited test supply during COVID-19 surges. To date, a rapid, accessible, and affordable testing method has not been deployed yet.

Emerging Artificial Intelligence (AI) technologies have shown the ability to create fast, affordable, and accessible solutions. There is increasing evidence that machine learning and deep learning methods can analyze cough sounds of infected patients and predict COVID-19 \cite{pja2021}. Multiple research groups have been gathering sound recordings for COVID-19 patients of all ages, in various settings, symptomatic and asymptomatic, and at different periods relative to symptom onset. These allow the AI algorithms to learn audio characteristics of COVID-19 illness in patients with various demographic and medical conditions. A potential, purely digital COVID-19 detection method would allow for a smartphone-based rapid, equitable COVID-19 test with minimal infection risk, economic burden, and supply chain issues - all helpful factors to control COVID-19 spread. Given the importance, there is an urgent need to explore and develop machine learning-based digital COVID-19 tests.

\section{Related Works}
\label{igw}

Studies reveal that cough samples can be used for diagnosing respiratory syndromes such as pneumonia, pulmonary diseases, and asthma \cite{thorpe}\cite{song}\cite{infante}. Recently, researchers have explored the possibility of screening COVID-19 using cough signals. In \cite{brown}, the authors have demonstrated that a binary machine learning classifier is sufficient to identify COVID-19 cough and breath signals from healthy individuals or asthmatic patients. In \cite{lag}, a framework for AI speech processing that pre-screens cough to identify COVID by leveraging acoustic bio-marker as features. In this study, 96 cough samples are collected from users with bronchitis and 136 with pertussis in addition to the COVID positive and healthy cough samples. Moreover, this study uses three independent, parallel classifiers, namely, a deep transfer learning-based multi-class classifier (DTL-MC), a classical machine learning-based multi-class classifier (CML-MC), and a deep transfer learning-based binary class classifier (DTL-BC). These three independent classifiers are used due to the insufficient data available for learning. Lastly, handcrafted features like spectral centroid (SC), spectral roll-off (SR), zero-crossing rate (ZCR), MFCC, delta-delta MFCC can be used to train long short-term memory (LSTM) network models as shown in \cite{hassan}. Classification on voice samples demonstrated the least accuracy among other acoustic features like cough and breath.

Mouawad et al. utilize various statistical measures to detect COVID-19 from cough and the sustained vowel sound \lq ah' \cite{Mouawad2021}. Best recurrence structures were extracted using variable Markov Oracle as a pre-processing stage. The DiCOVA challenge was initiated to encourage researchers in COVID-19 diagnosis using acoustics \cite{dicova}. Among the two datasets used in the DiCOVA challenge, one focuses only on cough sounds, and the other contains a collection of audio recordings, including counting numbers, phonation of sustained vowels, and breath. 11 most common symptoms combined with an established feature set of 384 features, obtained by applying 12 functions on the 16 frame-level descriptors and the respective delta coefficients, are classified using support vector machine (SVM) with a linear kernel in \cite{han}. In the first experiment, feature level fusion is implemented by concatenating the audio and the symptom features into a single matrix. In the second experiment, decision level fusion is explored by finding the maximum probability among two models trained independently, which is reported to yield superior performance. 

The evaluation in \cite{pahar} reveals that among the popular machine learning approaches like SVM, k-nearest neighbor (kNN), logistic regression (LR), multi-layer perceptron (MLP), ResNet50, and LSTM, the ResNet50 generated the highest AUC (0.976) on the Coswara dataset. In \cite{erdogan}, feature extraction using the traditional and deep learning approaches are compared through empirical mode decomposition (EMD) and discrete wavelet transform (DWT), where ReliefF is utilized as a feature selection technique. It is concluded that high performance is achieved when DWT features are combined with traditional ML approaches such as SVM. In \cite{despotovic}, experiments are conducted to identify the most informative acoustic features as a baseline instead of the handcrafted features or MFCC. It is also demonstrated that features can be learned from limited data resources using the wavelet scattering transform. 

In \cite{tena}, cough sounds are detected from the raw audio files using YAMNet, and the discriminatory features are identified by analyzing the time-frequency representation (TFR) of the Choi-Williams distribution. Recursive feature elimination is used to select useful features based on the feature importance measure. Several models were analyzed, and RF is reported to have superior performance in differentiating COVID positive and negative samples.

In table \ref{tab:table}, we have summarized the models used, their dataset, and the performance metrics of various state-of-the-art algorithms for COVID detection from cough samples. From the literature survey, it is evident that most researchers have exploited pre-trained models for determining COVID from cough due to the lack of large datasets.
\begin{table*}
\centering
\caption{{Dataset and Performance of State-of-the-art Algorithms}}
\begin{tabular}{lp{4cm}p{2cm}p{2cm}p{2cm}p{1.5cm}}
\hline
Reference & Model & \multicolumn{2}{c}{No. of Samples} & Metric Reported & Dataset Availability\\ \cline{3-4}
& & Covid-19 Positive & Covid-19 Negative & \\ \hline \hline \\
\cite{brown} & PCA reduced handcrafted \& VGGish features with binary classifier & 141 & 298 & AUC: 80 & On-request    \\

 \cite{lag} & MFCC with one Poisson biomarker layer and 3 pre-trained ResNet50 in parallel & 2660 & 2660 & \vtop{\hbox{\strut Sensitivity: 98.5} \hbox{\strut Specificity: 94.2} \hbox{\strut AUC: 97}} & Public    \\                       

\cite{hassan} & Handcrafted features like SC, SR, ZCR, MFCC and delta-delta are classified using LSTM & 60 & 180 & \vtop{\hbox{\strut F1-score: 97.9} \hbox{\strut AUC: 97.4} \hbox{\strut Accuracy: 97}} & Private \\

\cite{Mouawad2021} & MFCC with repetition structure classified using XGBoost & 32 & 1895 & \vtop{\hbox{\strut Accuracy: 97} \hbox{\strut AUC: 84}} & Private \\

\cite{dicova} & MFCC, Delta and Delta-delta coefficients were used as features to train LR, MLP and Random Forest (RF) classifiers & 62 & 380 & \vtop{\hbox{\strut LR-AUC: 66.95} \hbox{\strut MLP-AUC: 68.54} \hbox{\strut RF-AUC: 70.69}} & Public \\

\cite{han} & 11 Covid related symptoms with 384 features obtained from the voice samples are classified using SVM and decision-level fusion & 326 & 502 & \vtop{\hbox{\strut Sensitivity: 68} \hbox{\strut Specificity: 82} \hbox{\strut AUC: 79} \hbox{\strut Accuracy: 79}} & Private \\

\cite{pahar} & MFCC, MFCC velocity and acceleration along with log frame energies, ZCR and kurtosis are used as features with ResNet50 & 110 & 1105 & \vtop{\hbox{\strut AUC: 98} \hbox{\strut Accuracy: 95.33} \hbox{\strut Specificity: 98} \hbox{\strut Sensitivity: 93}} & Public \\

\cite{erdogan} & Intrinsic mode functions using EMD together with DWT features, ReliefF feature selection and SVM & 595 & 592 & \vtop{\hbox {\strut Specificity: 97.3}\hbox{\strut Accuracy: 98.4}\hbox{\strut F1-score: 98.6}} & Public (Virufy) \\
\cite{despotovic} & Wavelet scattering features with Boosting classifier & 84 & 1019 & \vtop{\hbox{\strut Accuracy:88.52} \hbox{\strut Sensitivity:87.19}\hbox{\strut Specificity: 89.82}} & Private \\

\cite{tena} & YAMNet with TFR analysis & 346 & 346 & \vtop{\hbox{\strut Accuracy: 83.67} \hbox{\strut AUC: 93.56} \hbox{\strut Sensitivity: 89.58} \hbox{\strut Specificity: 71.58}} & Public \\ \hline
\end{tabular}
\label{tab:table}
\end{table*}

\section{Methods}

In this section of our study, we explain the resource
we acquired along with a description of the steps taken to extract
information from the cough samples before they were handled by our model
development team. Afterward, we explore the techniques that were used to
train each model.

\subsection{Dataset}

{\label{123021}}

Multiple publicly available datasets were combined and utilized to
ensure minimal bias during model training.~ Among the crowdsourced data
that were utilized were COUGHVID~\cite{Orlandic2021} and
Coswara~\cite{sharma2020} datasets.~As a publicly-available dataset of
global cough audio recordings, COUGHVID is one of the largest COVID-19
related cough datasets. COUGHVID includes a total of 20,072 cough
recordings labeled as positive COVID-19, symptomatic COVID-19 negative,
and asymptomatic COVID-19 negative, along with other clinical
information and metadata~\cite{Orlandic2021}. The dataset contains
samples from a wide array of ages, genders, pre-existing respiratory
conditions, and geographic locations - all of which are useful data to
consider when training an unbiased deep learning model. Similarly, the
publicly available Coswara dataset was crowdsourced from participants
around the globe utilizing a smartphone-based data collection
platform~\cite{sharma2020}. Virufy processed the Coswara dataset and labeled over 2000 coughs
resulting in the diverse demographics of samples, with 1839 samples
as COVID-19 negative (``healthy'') and 411 samples as
COVID-19 positive (``infected''). Furthermore, to increase the usefulness of AI systems like ours in hospital settings, we sought to train
our model on clinically gathered datasets. Subsequently, we utilized the
IATOS dataset, which is a publicly available dataset gathered by the
city government of~ Buenos Aires, Argentina, using WhatsApp chatbots.
This validated dataset includes 2196 positive and 2199 negative patients
in 11 hospital facilities where RT-PCR studies were carried out on
patients suspected of COVID and 14 out-of-hospital isolation units for
patients with confirmed COVID mild cases~\cite{pizzo2021}.~

\begin{figure}[h!]
\begin{center}
\includegraphics[width=1\columnwidth]{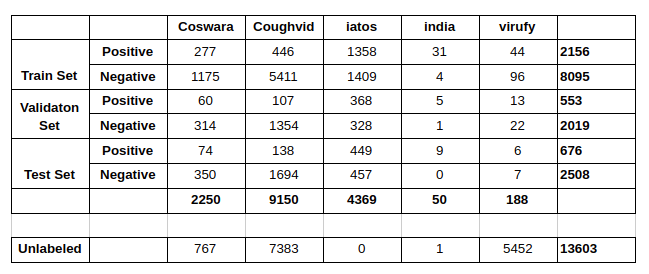}
\caption{{The train, validation and test sets along with unlabeled data set sizes
{\label{926439}}%
}}
\end{center}
\end{figure}

In addition to the aforementioned datasets, we also gathered and
utilized our own crowdsourced and clinical datasets. Virufy's
crowdsource dataset was collected using our study app
(\url{https://virufy.org/study})~which requests users to input
demographic information, such as age, gender, and comorbidities, along
with their PCR test results. This app is publicly advertised on our
website and as well as through press releases via Forbes, Yahoo, and
various media outlets in South America and India. Additionally, we took
sufficient measures to protect patient's privacy with our informed
consent forms and privacy policy
(\url{https://virufy.org/privacy\_policy/}). For Virufy's clinical
dataset, in recognizing that the majority of identified COVID-19 positive
patients present at clinics for testing, we developed a protocol for
data collection at clinics. After extensive ethical review and
scientific oversight, we achieved IRB approval for our data collection
method. In this method, patients presenting for COVID tests are first
requested to input demographic information, such as age, gender, and
comorbidities, into our app (\url{https://virufy.org/study}). After
doing so, they are requested to wear a surgical or cloth mask and cough
intentionally into the smartphone. Finally, when test results are
available, they are reported in the app and linked to the patient's
cough. This protocol increases the likelihood that our data quality is high and
clinically valid due to oversight by hospital staff and the IRB
committee. This oversight also helps to establish acceptance for use in
scientific research and clinical validation
studies.~Figure~{\ref{926439}} shows the labeled data
size for training, validation, and test sets as well as the unlabeled
data size that is used during unsupervised learning.\selectlanguage{english}

\subsection{Audio Preprocessing}
To provide readily usable data for model development, the cough audio data goes through two main stages: standardization and preprocessing. Data standardization repackages the raw data including, the data labels, naming conventions, and file types according to Virufy's internal Common Data Format (CDF). The preprocessing stage takes the standardized data to filter and prepare usable samples by detecting background noise and removing any inadequate data, such as samples with no coughs detected from it. The initial standardization and preprocessing scripts were in Colab notebooks and were run manually and separately tested for each dataset source. To make this more streamlined, we moved most of the code to one centralized GitHub repository (\href{http://}{https://github.com/virufy}). We automated much of the data processing pipeline and its corresponding testing, in addition to making improvements to cough detection and computation time.

\subsubsection{Standardization}
The dataset standardization was originally performed through individual Colab notebooks by data source; for example, FluSense data was processed separately from Coughvid data. Colab notebooks may require someone to manually run the preprocessing notebook, authenticate access to AWS S3 to upload the standardized data and check that the program finished running. This placed a significant burden on model developers, and the underlying structure made it difficult to keep track of what changes were made to which notebook. To provide a uniform, centralized infrastructure for standardization, we moved many of the notebooks to one Github repository and wrote Python scripts that automated the standardization including the authentication and upload. The repository provided an organized structure that allows us to make any additional changes transparently.

\subsubsection{Preprocessing}
Given that audio files from different datasets can have different numbers of audio channels and sampling rates, the audio files are first converted to mono and resampled at a 16 kHz sampling rate. A series of checks are applied to the standardized audio to remove low-quality audio files. The following checks are conducted:
\paragraph{Volume Detection}
The volume of the audio file needs to be sufficient as it would be difficult to extract meaningful data even if amplification is used. The maximum amplitude of the audio file is calculated to detect the volume of the audio file. 
\paragraph{Clipping Detection}
Clipping indicates that some samples in the audio file are missing as the amplitude of the samples was truncated. The ratio of clipping in the audio file is calculated by analyzing peaks and their flatness. 
\paragraph{Cough Detection}
A pre-trained CNN model based on ResNet-18 architecture provided in \cite{bagad2020cough} is used to determine if the audio file contains a cough. The CNN model outputs a probability of whether a cough was detected in the audio file.
\paragraph{Cough Segmentation}
The regions that contain coughs in the audio file are determined through the segmentation process described in \cite{2021}. Unlike the other checks that are conducted, the cough segmentation algorithm uses audio files that are converted to mono and resampled at 44.1 kHz as inputs. The segmentation algorithm first removes background noise through a combination of low pass and high pass filters and resamples the filtered audio with a 4.41 kHz sampling rate. 

\paragraph{Background Noise Detection}
Loud background noise can obscure the audio characteristics of the signal in interest. To evaluate the loudness of the background noise in the audio file, the ratio between the max power of cough segments and non-cough segments is calculated. The cough segments and non-cough segments are determined in the cough segmentation check.

The outputs from the aforementioned checks are used as thresholds to determine low-quality audio files. Audio files that do not meet the thresholds are removed and are not used in subsequent steps, such as feature extraction, model training, and inference.

\subsection{Support Vector Machine}
One of the effective supervised learning methods in machine learning is support vector machine (SVM) \cite{vapnik}. The classes are separated from each other by a hyperplane passing through the closer elements. SVMs perform well when the data is linearly separable. However,  kernel functions can be used to non-linearly separate the data when it is not linearly separable. In our experiments, we have utilized the radial basis function for the SVM kernel. To map the SVM outputs to probabilities within the [0, 1] range, the SVM outputs are fed into logistic regression. Platt Scaling is used to transform the SVM output to probability values \cite{platt1999}.

Short-time fourier transform (STFT) and mel-frequency cepstral coefficients (MFCC) \cite{rabiner} have been used by several researchers for classifying the Covid-19 cough or voice samples \cite{lag}\cite{hassan}\cite{Mouawad2021}.    
The handcrafted features such as spectral centroid, spectral roll-off, root mean square, Mel-frequency cepstral coefficients (MFCC), delta-MFCC, and double delta MFCC are used for classification using the support vector machine (SVM). Mean, standard deviation, maximum and minimum of the above-mentioned features are computed wrt time and are then normalized before being used by the SVM.

\subsubsection{Spectral Centroid}
Spectral centroid (SC) measures the change in frequency and phase content over time and it defines the shape of the magnitude spectrogram of STFT by determining its center of gravity from the normalized frame \cite{nam}. 

\subsubsection{Spectral Roll-off}
Spectral roll-off (SR) measures the skewness of the spectrogram by determining the central frequency of the spectrogram bin at which 25\% and 75\% of the magnitude are concentrated \cite{scheirer}.

\subsubsection{Root Mean Square of the Spectrogram}
The loudness of the signal can be determined using the root mean square (RMS) of the spectrogram. Lower RMS values are a result of high signal energy in a block of the spectrogram \cite{mulimani}. 

\subsubsection{Mel-frequency Cepstral Coefficients}
The fast Fourier transform (FFT) bins from the log-amplitude of the magnitude spectrum are grouped and smoothed according to the Mel-scale. Finally, discrete cosine transform is applied to de-correlate the feature vector. It is found that 13 MFCCs are sufficient to represent speech, voice, or cough samples \cite{tzanetakis}.

MFCCs are given by

\begin{equation}
C\left(m\right)=\sum_{n=1}^N \log\, Y(n)cos\left(\frac{n\pi }{M}\left(m-\frac{1}{2}\right)\right)
\end{equation}

when m denotes the index of the MFCC and $Y(n)$ denotes the output of the n-channel of the filter bank.

\subsubsection{$\Delta$ and $\Delta ^2$ MFCC}
The first-order derivative of MFCC is given by \cite{kumar}

\begin{equation}
D[n] = C[n+r]-C[n-r]
\end{equation}

where $r$ generally takes 2 or 3 as a value. $\Delta ^2$ MFCC is obtained by determining the derivative of the subsequent delta cepstral features. 

\subsection{Self-Supervised Learning with
Transformers}

{\label{767472}}

In our work, we also used the Self Supervised Learning (SSL) method with
the unlabeled data to pre-train the network and we added labeled data to
the pretraining dataset to learn better representations. In
addition, we utilized an auxiliary task of predicting masked input while
reducing the reconstruction loss. To the best of our knowledge, the work
that is closest to our approach is that of Xue et
al.~\cite{salim2021}. In contrast,~\cite{salim2021} used masking to
prevent overfitting and used contrastive loss. Both approaches employ
popular transformer architectures, but we used the spectrogram image as
input to the transformer network and masked the input spectrogram both
in the time and frequency domain. In contrast, Xue et al.
\cite{salim2021} used MFCC and applied masking in the time domain
only. Our work used the s3prl~\cite{toolkit} library for pretraining the
upstream network and training the downstream network. More specifically,
we have adapted upstream TERA~\cite{tera}~ and modified downstream
Speaker Classification parts of the library for our Covid-19
classification task from audio samples. We did not employ any
pre-trained network weights. Both upstream and downstream models are
trained from scratch by using the training set described in
Figure~{\ref{926439}}.

\subsubsection{Architecture}

{\label{736602}}

\begin{figure*}[h!]
\begin{center}
\includegraphics[width=0.84\textwidth]{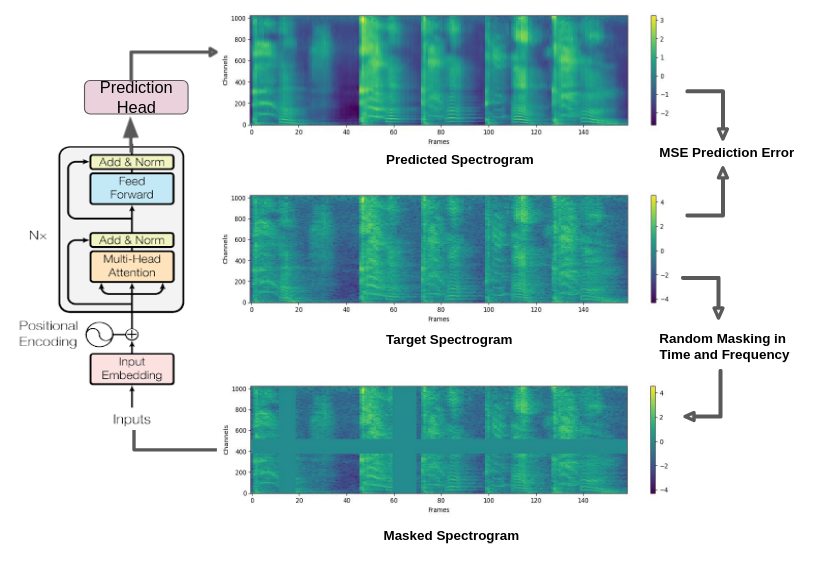}
\caption{{Upstream Training with Transformers
{\label{774778}}%
}}
\end{center}
\end{figure*}

Our model uses the transformer encoder architecture to extract
representations from cough audio signals. The transformer encoder
network has three layers, each with a hidden layer size of 768, 12
attention heads, and a feed-forward network of 3072. Transformer encoder
layers learn representation by masked signal prediction task as shown in
Figure~{\ref{774778}}. The masked signal prediction
task could be described as follows. The input cough audio is transformed
to a spectrogram image and then randomly masked in frequency and time
domain as described in~{\ref{473686}}. This masked
spectrogram image is fed into the transformer encoder layers.
Transformer encoder layers learn representations while the prediction
network following the transformer encoder layers tries to reconstruct
the original spectrogram of the cough audio. The prediction network
consists of 2 feed-forward layers with a hidden size of 768.~ The
reconstruction loss is backpropagated through the prediction network and
transformer encoder layers. This process is also called upstream
training. Once upstream training is complete, the prediction network is
removed, the upstream model is frozen, and representations from the
transformer encoder are fed into the downstream model. The upstream
model is based on the TERA~\cite{tera} implementation of
the s3prl~\cite{toolkit} library.

The downstream model consists of three linear~ layers followed by
normalization and non-linear activation layers, a mean pooling layer, and
a classification layer. The first linear layer converts the hidden size
768 of encoder output to 512. The following two linear layers both have
size 512. Next, the mean pooling layer calculates the mean of the
representation across the time axis, ending up with a 512-dimensional
vector. Finally, the classification layer uses a 512-dimensional vector
to classify cough audio as either COVID or not COVID. The downstream
model training is done by backpropagating the classification error
through the downstream model. The downstream model is adapted and
modified from the Speaker Classification task of the s3prl~\cite{toolkit}
library.

\subsubsection{Input}

{\label{473686}}

Virufy preprocessing library detects cough boundaries, and cough
boundary metadata is given to the model alongside the input audio. The
audio signal is then converted to cough-only audio by removing the
silent parts in the audio signal. The cough-only audio signal is then
converted into a spectrogram image using librosa~\cite{librosa}~Short-Time
Fourier Transform (STFT) function with n\_freq=2048, win\_length=640,
and hop\_length=320 at 16KHz audio sample rate. The logarithm of the
spectrogram is finally taken. With these values of STFT function, each
40~{ms} of the audio signal is converted into 1025 frequency bins with a
shift of 20 ms along the time axis. The spectrogram time axis varies
depending on the input audio length.~ This variation is handled by first
padding the audio signal with zeros for batch processing and then
masking the input for later stages of input processing.

During upstream training, the log spectrogram is masked both in time and
frequency bands before sending it to the transformer encoder. The amount
of mask in the time domain is set as 0.15, while the maximum amount of
masked frequency bands is 0.20. Additionally, 0.10 of time, a Gaussian
noise of mean 0 and variance 0.2 is added to the spectrogram image.~

Masking is not applied to log spectrogram during downstream training.
The log spectrogram is input to the frozen upstream model, and the
transformer encoder outputs the input representation. This
representation is fed into the downstream model.\selectlanguage{english}

\subsubsection{Upstream Training}

{\label{150043}}

During upstream training, an auxiliary task of spectrogram prediction
from masked input spectrogram is used to learn the representation of the
input. This task enables self-supervised learning using the unlabeled
data in addition to labeled data. As a result, the transformer encoder
can learn better representations for cough audio by using more data.~

We trained the upstream model using $\approx$13K unlabeled and
$\approx$10K labeled cough audio samples (see Figure
{\ref{926439}}) using a batch size of 64, with a
maximum learning rate of 1.e-4 over 400K steps. The learning rate warms
up from zero to a maximum value over the first 28K steps and decays
linearly to zero afterward. We used MSE as a loss function during
reconstruction and the AdamW~ optimizer~\cite{loshchilov2017decoupled} to update the
upstream model parameters.

\subsubsection{Downstream Training}

{\label{228544}}

Once upstream model training is complete, we freeze the upstream model
parameters at the end of the 400K steps and start with downstream model
training. During downstream training, we only use labeled data. We use
the same $\approx$10K labeled data used during upstream
training. We set batch size as 8, learning rate as 1.e-4, and used AdamW
optimizer. Due to the imbalance in positive and negative classes, we used an
upsampling strategy for positive classes with a ratio of 5:1. Compared
to upstream, downstream training takes much fewer steps to train. We set
the steps as 6K with an initial 1K step of warming up. After every 300
training steps, we evaluated the network with $\approx$ 2600
validation samples. We picked the downstream parameters that give the
maximum AUC.~ We observed that downstream training converges after 3-4
epochs.

\subsection{Convolution Neural
Network}

{\label{464945}}

This model is a continuation of the work presented by Rasmussen et al
(2021) at the Applied Machine Learning and Data Analytics conference at
the end of 2021. This manuscript is currently in preparation for
submission to a peer-reviewed journal~\cite{rasmussen}.~ This model
was inspired by modern convolutional networks with a few key
differences. First, there are no Batch Normalization (BN) layers.
Generally, BN is used in residual networks (ResNets) for object
classification and detection from RGB images. We have not found BN
effective when we are working with monochrome sonographs. The second
difference is that this architecture does not use skip connections.
Traditional ResNets use skip connections due to extreme depth, and skip
connections help to alleviate the vanishing gradient problem. This
architecture does not use skip connections since it is shallow compared
to modern ResNets. Minor modifications have been made to the model
compared to what will be published in the conference paper. The
differences are how the sonographs have been created and the extension
of the CNN's repeated layers, along with a global average pooling layer
instead of flattening the features before being passed to the dense
layers.

\subsubsection{Balancing the Dataset}

{\label{553738}}

Before creating the sonographs, a direct 4:1 ratio of up-sampling of all
positives from each training dataset is performed. This up-sampling
allowed the combination of the datasets to be approximately balanced
regarding the ratio of positives to negatives. After up-sampling all
positives from each dataset, all samples, including the negatives, were
randomly augmented using the library
``Audiomentations''~\cite{learning}. The augments that were performed include
shifting the audio time-wise and adding a Gaussian single-to-noise ratio.
The values used for the ``Shift'' function were: a min\_fraction of -.5,
a max fraction of .5, and a p-value of 1. These values ensured that all
of the new samples created were shifted randomly time-wise and guaranteed
all of the new samples were augmented. The values used for the
``AddGaussianNoise'' function were: a min\_SNR of .25, a max\_SNR of .9,
and a p-value of .5. This step increased the likelihood that approximately half of all
the new samples created had a Gaussian signal-to-noise ratio added to them
in between the specified interval. These steps created a resulting
training set that had one set of normal negatives, one set of augmented
negatives, four sets of normal positives, and four sets of augmented
positives.

\subsubsection{Feature Extraction}

{\label{321664}}

For the creation of each sonograph from the samples described above, the
library Librosa was used~\cite{081rc2}. Once the audio was loaded with the library, the audio
was condensed using the cough segments that were extracted by the
pre-processing step described in Section 3.1 of this paper. Afterward,
regardless of how many seconds of audio was extracted by the cough
segmenter, all samples were fit into an approximate 4 second window. If
there were less than 4-seconds of audio extracted per sample, zeros were
padded at the end of the audio. If more than 4 seconds were extracted
per sample, the audio was cut off. This previous step allowed all
sonographs created to have a standard size and prevented the need to
resize the sonographs, which dilates it and negatively affects the data
quality. Once all samples were put into a standardized data array,
augmentations described above were performed and Librosa extracted the
MFCC features of the audio. Sixty-four MFCCs were used, along with a hop
length of 256 and an n\_fft window size of 512. These values were used
as opposed to the default values because they allowed us to ``zoom in''
on the temporal features of the audio as opposed to concentrating on the
frequency resolution of the sonograph. The resulting sonograph was a 64
by 256-pixel MFCC representation of the four-second condensed audio. ~

\subsubsection{Architecture}

{\label{822160}}

Our study initially passed the input described above to a convolution
layer with a stride value of 1 by 1, 32 filters, and a dimension of 7 by
7 to look for the most general features. Next, this output was passed to
another convolution layer with a stride value of 1 by 1, 64 filters, and 5 $\times$ 5 dimension, to look for slightly less general features with
roughly 32-fold the number of nodes. Afterward, the previous layer's
output was passed to a 2 by 2 max-pooling layer with a stride of 2 by 2
to halve the spatial data vertically and horizontally for the next
layer. This layer is followed by another sequence of convolution and
pooling layers. In this case, the convolution layer consisted of a
stride value of 1 by 1, 256 filters, and the dimension of 3 $\times$ 3, and the
max-pooling layer was the same as the previous. These previous two
layers were repeated twice and helped derive more specific features from
the previous layers. These repeated layers were followed by a final
convolution layer with a stride value of 1 by 1, 512 filters, and
dimension of 3 $\times$ 3, which helps find only the smallest features from
the previous layers with the largest amount of nodes in the network. The
final output of the last convolutional layer was then passed to a Global
Average Pooling layer which condenses the output to the 256-dimensional
dense layer. The output from this dense layer was routed to the final
output layer, classifying the image as a zero or one. All layers use a
ReLU activation function and a dropout rate of 0.2, while the final
layer uses a sigmoid activation with no dropout for the final output.
Table~\protect\hyperlink{author-label-layers}{2}~has been provided below
to show the above layout of the proposed CNN architecture. ~

\subsubsection{Training
Hyper-parameters}

{\label{702960}}

The model was trained with a learning rate of .0001 for 100 epochs and a
batch size consisting of 16. Model checkpoints were used based on the
best AUC achieved on the validation set to prevent overfitting. A binary
cross entropy-based loss function was used along with the Adamax
optimizer. The details of the number of generated parameters for the
different layers are presented in Table {\ref{layers}}.\selectlanguage{english}
\begin{table*}[tbp]
\center
\caption{{Hyper-parameters used in different layers of the proposed CNN.}}
\label{layers}
\begin{tabular}{llllll}
\hline 
Layer &Strides\hspace{.75mm} &Window Size \hspace{0.5mm} &Filters \hspace{.5mm} & Parameters \hspace{.5mm} & Output\\ \hline \hline
Convolution 1 \hspace{1mm} & $1\times1$     &$7\times 7$         &32     & 1,600        & (64x256x32) \\
Convolution 2              &$1\times1$      &$5\times5$         &64     & 51,264       & (64x256x64) \\
Max Pooling                &$2\times2$      &$2\times2$         &N/A    & N/A          & (32x128x64)  \\
Convolution 3              &$1\times1$      &$3\times3$         &256    & 147,712      & (32x128x256) \\
Max Pooling                &$2\times2$     &$2\times2$         &N/A    & N/A          & (16x64x256) \\
Convolution 4              &$1\times1$      &$3\times3$         &256    & 590,080      & (16x64x256) \\
Max Pooling                &$2\times2$     &$2\times2$         &N/A    & N/A          & (8x32x256)  \\
Convolution 5              &$1\times1$      &$3\times3$         &512    & 590,080      & (8x32x256)  \\
Max Pooling                &$2\times2$     &$2\times2$         &N/A    & N/A          & (4x16x256)  \\
Convolution 6              &$1\times1$      &$3\times3$         &512    & 1,180,160      & (4x16x512)  \\
GlobalAveragePool         &N/A     &N/A         &N/A    & N/A          & (512)     \\
Dense 1                    &N/A     &N/A         &256    & 131328     & (256)       \\
Dense 2                    &N/A     &N/A         &1      & 257          & 0 or 1      \\ \hline
\end{tabular}
\end{table*} 

\section{Results}

In this section of our study, we provide a comparative analysis of the
models presented in our work. We explain the
validation results, along with{ the AUC for~}our test set. Future work
is planned to further modify and improve our models, along with
collecting more data.

\subsection{Validation Results}

{\label{105286}}

To compare our deep learning models to a simple baseline architecture,
we used our shallow learning SVM model. SVM model achieved a specificity
of 0.96, a sensitivity of 0.14, and an AUC of 0.75. These numbers are
heavily skewed toward specificity, and could be alleviated by modifying
the output thresholds, however that would most likely come at a cost to
other metrics to improve the sensitivity. The self supervised learning
model achieved much more balanced results with a specificity of{ 0.51},
a sensitivity of 0.89, and an AUC of 0.807 on the validation data. Also,
the CNN performed comparably with a specificity of 0.74, a sensitivity
of 0.77, and an AUC of 0.802.~Note that sensitivity, specificity, and
AUC values are reported at an output threshold of 0.5.~
Table~{\ref{tab2}} has been provided later in the~paper
to compare all performance metrics collected. ~

\par\null

\subsection{Different Samples Sizes}

{\label{796763}}

To provide analysis on how the amount of data can affect a
model's performance, we show the self supervised and CNN model
performance with different dataset sizes. To do this, we have
provided the results on the same validation set while decreasing the
number of samples used for training by a factor of .8, .6, .4, and .2
using random sampling of the dataset. Although there is only a slight
performance degradation, we can see there is a general downward trend of
performance as the sample sizes decrease, also the performance metrics
of the models are more variable, with some of the metrics skewed towards
specificity or sensitivity. Table~{\ref{tab2}} will
compile these results as well.

\subsection{Test Results}

{\label{975063}}

{ For the test results, we have under reported the
performance metrics of both the self supervised learning and CNN models
to limit any bias in our model for future work. Currently, we
are planning on modifying all the models currently being researched at
Virufy, adding more samples to the many datasets we have already
collected, and publishing a new paper with more complete performance
metrics on the enlarged dataset. As such, we have decided to only report
the AUC for the test data. The self supervised learning model
achieved~}an AUC of 0.791.~Also, the CNN performed comparably with an
AUC of 0.775.~ These performance metrics are also tuned to the output
threshold of .5 to delineate positives and negatives and have been
compiled in table {\ref{tab2}}.\selectlanguage{english}
\begin{figure}[h!]
\begin{center}
\includegraphics[width=0.85\columnwidth]{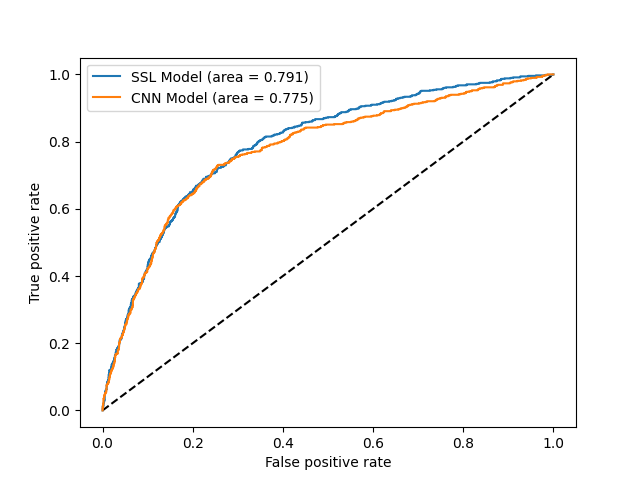}
\caption{{ROC-AUC curve of both models on the test data.
{\label{850421}}%
}}
\end{center}
\end{figure}\selectlanguage{english}
\begin{table*}[tbp]
\center
\caption{{Covid-19 screening performance in Accuracy (ACC), Area Under the Curve (AUC), Sensitivity (SEN) and Specificity (SPEC)}}
\label{tab2}
\begin{tabular}{llcccc}
\hline 
&& \multicolumn{4}{c}{\bf Performance}\\
               \textbf{Model} & \textbf{Cough samples} & \textbf{ACC} &\textbf{AUC} &\textbf{SEN}  &\textbf{SPEC}              \\ 
               \hline
               \hline
\textbf SVM
    & 100\% Train Vs Validation     & 0.79       & 0.75      & 0.14    &  0.96\\

\hline

\textbf Self Supervised
    & 100\% Train Vs Validation     & 0.59       & 0.807      & 0.89    &  0.51\\
    
\textbf Self Supervised
    & 80\% Train Vs Validation   & 0.57       & 0.797      & 0.89    &  0.48\\
    
\textbf Self Supervised
    & 60\% Train Vs Validation   & 0.66       & 0.791      & 0.82    &  0.61\\
    
\textbf Self Supervised
    & 40\% Train Vs Validation    & 0.52       & 0.782      & 0.90    &  0.42\\
    
\textbf Self Supervised
    & 20\% Train Vs Validation    & 0.69       & 0.763      & 0.75    &  0.67\\
    
\textbf Self Supervised
    & 100\% Train Vs Test &  N/A      & 0.791      & N/A    & N/A\\
\hline 
\textbf CNN
    & 100\% Train Vs Validation   &  0.75      & 0.802      & 0.77    & 0.74\\

\textbf CNN
    & 80\% Train Vs Validation &  0.73      & 0.792      & 0.78    & 0.71\\
    
\textbf CNN
    & 60\% Train Vs Validation &  0.67      & 0.795      & 0.82    & 0.63\\
    
\textbf CNN
    & 40\% Train Vs Validation &  0.66      & 0.789      & 0.81    & 0.61\\
    
\textbf CNN
    & 20\% Train Vs Validation &  0.73      & 0.787      & 0.75    & 0.72\\
    
\textbf CNN
    & All Train Vs Test &  N/A      & 0.775      & N/A    & N/A\\

\hline

\end{tabular}
\end{table*}

\section{Discussion}

In this study, we compile a dataset of approximately 30,000 high-quality
coughs including around 16,000 labeled coughs and about 3,400 COVID-19
positive coughs from globally crowdsourced and clinical sources. We also
show that the self-supervised, CNN, and SVM approaches achieve an~{AUC
of 0.807, 0.802, and 0.75 on the validation set respectively, and that
the self-supervised and CNN approaches achieve an AUC of 0.791 and 0.775
on the test set respectively}.

Our best performing network, the SSL approach, had a sensitivity of 0.87
at a specificity of 0.49. Although the metrics are lower when compared
to the rapid antigen (sensitivity \textgreater{} 80\%, specificity
\textgreater{} 97\%;~\cite{a2021}) and the rt-PCR tests, our
approach has the advantage of being easily repeatable. Multiple testing
can increase the accuracy of the results. Furthermore, a cough-based
approach has the advantage of being readily accessible on mobile phones
and requiring less time and cost for results than any other form of
testing.

This effort builds on previous works showing the experimental efficacy
of detecting COVID-19 from coughs~\cite{pja2021}. One of the main
issues preventing wider adoption of cough-based COVID tests has been the
persistent doubt about bias and generalizability~\cite{topol2020,schulle2021}. Our
study advances prior work in the field by applying deep learning
techniques to the largest aggregate evaluation dataset of rt-PCR tested,
high-quality, COVID-19 positive coughs paired with COVID-19 negative
coughs collected in similar settings. Furthermore, our datasets includes
crowdsourced coughs (Virufy, COUGHVID, Coswara) and PCR-verified
clinical ones (Virufy-India, IATOS). By using datasets from multiple
sources, we hope to ameliorate dataset-inherent bias. Additionally, our
results on varying the sample size show a trend of increasing
performance with increasing sample size, identifying the need to
continue collecting coughs to improve model performance.~

In this study, we compare the performance of a self-supervised learning
(SSL) and CNN approach to a baseline SVM model. As expected, we find
that the SVM, a simpler model, is not able to differentiate the cough
signals as well as the SSL and CNN models. We also compare the SSL
approach that takes advantage of unlabeled coughs with the more
traditional fully-supervised CNN approach. There is a small difference
between the models' AUCs, with the SSL approach having a 0.016 higher
AUC on the test set, however the SSL model is slightly skewed toward
sensitivity. Currently, it is unclear which model is better. There can
be multiple improvements made to both models with different types of
up-sampling techniques, hyperparameter tuning, output threshold tuning,
and over-fitting prevention techniques. There can also be improvements
to the pre-processing and feature extraction steps such as separating
the cough segments into individual samples and concatenating them for a
final prediction on a single audio file. Also, the type of data
collected by the Virufy team, and other teams around the world, can
affect model performance. Another possible next step is to experiment
with a fusion model incorporating predictions from the SSL-transformer,
CNN, SVM, and other approaches to improve prediction performance.~

There are several limitations that should be noted. Previous studies
have demonstrated a higher AUC and other metrics than the ones that we
show in our study. However, our study's performance is likely closer to
the real-world performance of such a test due to our larger and more
diverse training dataset. Furthermore, most academic papers regarding
COVID-19 classification with AI suffer from reproducibility issues,
along with the fact that a lot of papers have only a small validation set,
and no test set, which increases the possibility of biased models.
Another limitation is that the labels from our crowdsourced training
data (72.6\% of labeled dataset) are unverified. This unverified data
has incorporated error into our models and will affect the performance
in real world environments. Future work should be done to invest
resources in collecting more PCR-verified coughs. Lastly, the
crowdsourced audio data is prone to noise that may lead to algorithmic
prediction bias. We try to account for this issue by only including
high-quality data from all sources using an internally developed cough
screening tool. Furthermore, our inclusion of multiple data sources
should reduce the effect of background noise specific to datasets.

\section{Conclusion}

This study is a step forward toward an accurate, reliable, and
accessible audio-based COVID-19 diagnostic test, showing adequate
performance on a large and diverse dataset. In addition to improving
model performance, future work should focus on collecting high-volumes
of PCR-validated samples and assessing the model generalizability to new
datasets. Decision-making errors should be closely studied to understand
how prediction performance is affected by audio recording conditions,
patient symptoms, and the time point during the course of COVID-19
infection.

\section{Acknowledgements}

We are very grateful to Mary L. Dunne, M.D., Stanford University
Distinguished Career Institute Fellow, for their kind guidance with
respect to the medical implications of our research, and Rok Sosic,
Stanford University Senior AI Researcher, for his guidance on team
structure and academic collaboration. We appreciate Siddhi Hedge and
Shreya Sriram for their amazing enthusiasm and hard work in facilitating
clinical cough data collection from COVID-19 tested patients.
M. Pilanci is partially supported by the National Science Foundation, and the U.S. Army Research Office. This study was supported by the Amazon Web Services Diagnostic Development Initiative.
\bibliographystyle{IEEEtran}
\bibliography{biblio.bib%
}

\begin{thebibliography}{10}
\providecommand{\url}[1]{#1}
\csname url@samestyle\endcsname
\providecommand{\newblock}{\relax}
\providecommand{\bibinfo}[2]{#2}
\providecommand{\BIBentrySTDinterwordspacing}{\spaceskip=0pt\relax}
\providecommand{\BIBentryALTinterwordstretchfactor}{4}
\providecommand{\BIBentryALTinterwordspacing}{\spaceskip=\fontdimen2\font plus
\BIBentryALTinterwordstretchfactor\fontdimen3\font minus
  \fontdimen4\font\relax}
\providecommand{\BIBforeignlanguage}[2]{{%
\expandafter\ifx\csname l@#1\endcsname\relax
\typeout{** WARNING: IEEEtran.bst: No hyphenation pattern has been}%
\typeout{** loaded for the language `#1'. Using the pattern for}%
\typeout{** the default language instead.}%
\else
\language=\csname l@#1\endcsname
\fi
#2}}
\providecommand{\BIBdecl}{\relax}
\BIBdecl

\bibitem{dashboard}
\BIBentryALTinterwordspacing
``{WHO Coronavirus (COVID-19) Dashboard},'' https://covid19.who.int/, accessed
  on Mon, January 03, 2022. [Online]. Available: \url{https://covid19.who.int}
\BIBentrySTDinterwordspacing

\bibitem{wadhera2021}
R.~K.~W. Abraar~Karan, ``{Healthcare System Stress Due to Covid-19: Evading an
  Evolving Crisis},'' \emph{Journal of Hospital Medicine}, 2021.

\bibitem{Mahase2021}
E.~Mahase, ``{Covid-19: Do vaccines work against omicron-and other questions
  answered.}'' \emph{BMJ}, vol. 375, p. n3062, Dec 2021.

\bibitem{Ma2021}
Q.~Ma, J.~Liu, Q.~Liu, L.~Kang, R.~Liu, W.~Jing, Y.~Wu, and M.~Liu, ``{Global
  Percentage of Asymptomatic SARS-CoV-2 Infections Among the Tested Population
  and Individuals With Confirmed COVID-19 Diagnosis: A Systematic Review and
  Meta-analysis.}'' \emph{JAMA Netw Open}, vol.~4, p. e2137257, Dec 2021.

\bibitem{kruger2021}
L.~Krüger, A.~Tanuri, A.~Lindner, M.~Gaeddert, L.~Köppel, F.~Tobian,
  L.~Brümmer, J.~Klein, F.~Lainati, P.~Schnitzler, O.~Nikolai, F.~Mockenhaupt,
  J.~Seybold, V.~Corman, T.~Jones, C.~Drosten, C.~Gottschalk, S.~Weber,
  S.~Weber, O.~Ferreira, D.~Mariani, S.~N.~E. Dos, P.~C.~T. Pereira,
  R.~Galliez, D.~Faffe, I.~Leitão, S.~R.~C. Dos, T.~Frauches, K.~Nocchi,
  N.~Feitosa, S.~Ribeiro, N.~Pollock, B.~Knorr, A.~Welker, V.~M. de, J.~Sacks,
  S.~Ongarello, and C.~Denkinger, ``{Accuracy and ease-of-use of seven
  point-of-care SARS-CoV-2 antigen-detecting tests: A multi-centre clinical
  evaluation.}'' \emph{EBioMedicine}, vol.~75, p. 103774, Dec 2021.

\bibitem{Scohy2020}
A.~Scohy, A.~Anantharajah, M.~Bodéus, B.~Kabamba-Mukadi, A.~Verroken, and
  H.~Rodriguez-Villalobos, ``{Low performance of rapid antigen detection test
  as frontline testing for COVID-19 diagnosis.}'' \emph{J Clin Virol}, vol.
  129, p. 104455, Aug 2020.

\bibitem{pja2021}
K.~K. Lella and A.~PJA, ``{A literature review on COVID-19 disease diagnosis
  from respiratory sound data},'' \emph{AIMS Bioengineering}, vol. 8(2), 2021.

\bibitem{thorpe}
W.~Thorpe, M.~Kurver, G.~King, and C.~Salome, ``{Acoustic analysis of cough},''
  in \emph{The Seventh Australian and New Zealand Intelligent Information
  Systems Conference, 2001}, 2001, pp. 391--394.

\bibitem{song}
I.~Song, ``{Diagnosis of pneumonia from sounds collected using low cost cell
  phones},'' in \emph{2015 International Joint Conference on Neural Networks
  (IJCNN)}, 2015, pp. 1--8.

\bibitem{infante}
C.~Infante, D.~Chamberlain, R.~Fletcher, Y.~Thorat, and R.~Kodgule, ``{Use of
  cough sounds for diagnosis and screening of pulmonary disease},'' in
  \emph{2017 IEEE Global Humanitarian Technology Conference (GHTC)}, 2017, pp.
  1--10.

\bibitem{brown}
\BIBentryALTinterwordspacing
C.~Brown, J.~Chauhan, A.~Grammenos, J.~Han, A.~Hasthanasombat, D.~Spathis,
  T.~Xia, P.~Cicuta, and C.~Mascolo, ``{Exploring Automatic Diagnosis of
  COVID-19 from Crowdsourced Respiratory Sound Data},'' in \emph{Proceedings of
  the 26th ACM SIGKDD International Conference on Knowledge Discovery \& Data
  Mining}, ser. KDD '20.\hskip 1em plus 0.5em minus 0.4em\relax New York, NY,
  USA: Association for Computing Machinery, 2020, p. 3474–3484. [Online].
  Available: \url{https://doi.org/10.1145/3394486.3412865}
\BIBentrySTDinterwordspacing

\bibitem{lag}
J.~Laguarta, F.~Hueto, and B.~Subirana, ``{COVID-19 Artificial Intelligence
  Diagnosis Using Only Cough Recordings},'' \emph{IEEE Open Journal of
  Engineering in Medicine and Biology}, vol.~1, pp. 275--281, 2020.

\bibitem{hassan}
A.~Hassan, I.~Shahin, and M.~B. Alsabek, ``{COVID-19 Detection System using
  Recurrent Neural Networks},'' in \emph{2020 International Conference on
  Communications, Computing, Cybersecurity, and Informatics (CCCI)}, 2020, pp.
  1--5.

\bibitem{Mouawad2021}
P.~Mouawad, T.~Dubnov, and S.~Dubnov, ``{Robust Detection of COVID-19 in Cough
  Sounds: Using Recurrence Dy source = {SN Comput Sci}, authors = {Monamics and
  Variable Markov Model.}}, date = {2021},uawad, p and dubnov, t and dubnov,
  s,'' \emph{SN Comput Sci}, vol.~2, p.~34, 2021.

\bibitem{dicova}
A.~Muguli, L.~Pinto, N.~R., N.~Sharma, P.~Krishnan, P.~K. Ghosh, R.~Kumar,
  S.~Bhat, S.~R. Chetupalli, S.~Ganapathy, S.~Ramoji, and V.~Nanda, ``{DiCOVA
  Challenge: Dataset, task, and baseline system for COVID-19 diagnosis using
  acoustics},'' 2021.

\bibitem{han}
J.~Han, C.~Brown, J.~Chauhan, A.~Grammenos, A.~Hasthanasombat, D.~Spathis,
  T.~Xia, P.~Cicuta, and C.~Mascolo, ``{Exploring Automatic COVID-19 Diagnosis
  via Voice and Symptoms from Crowdsourced Data},'' in \emph{ICASSP 2021 - 2021
  IEEE International Conference on Acoustics, Speech and Signal Processing
  (ICASSP)}, 2021, pp. 8328--8332.

\bibitem{pahar}
\BIBentryALTinterwordspacing
M.~Pahar, M.~Klopper, R.~Warren, and T.~Niesler, ``{COVID-19 cough
  classification using machine learning and global smartphone recordings},''
  \emph{Computers in Biology and Medicine}, vol. 135, p. 104572, 2021.
  [Online]. Available:
  \url{https://www.sciencedirect.com/science/article/pii/S0010482521003668}
\BIBentrySTDinterwordspacing

\bibitem{erdogan}
\BIBentryALTinterwordspacing
Y.~E. Erdoğan and A.~Narin, ``{COVID-19 detection with traditional and deep
  features on cough acoustic signals},'' \emph{Computers in Biology and
  Medicine}, vol. 136, p. 104765, 2021. [Online]. Available:
  \url{https://www.sciencedirect.com/science/article/pii/S001048252100559X}
\BIBentrySTDinterwordspacing

\bibitem{despotovic}
\BIBentryALTinterwordspacing
V.~Despotovic, M.~Ismael, M.~Cornil, R.~M. Call, and G.~Fagherazzi,
  ``{Detection of COVID-19 from voice, cough and breathing patterns: Dataset
  and preliminary results},'' \emph{Computers in Biology and Medicine}, vol.
  138, p. 104944, 2021. [Online]. Available:
  \url{https://www.sciencedirect.com/science/article/pii/S0010482521007381}
\BIBentrySTDinterwordspacing

\bibitem{tena}
\BIBentryALTinterwordspacing
A.~Tena, F.~Clarià, and F.~Solsona, ``{Automated detection of COVID-19
  cough},'' \emph{Biomedical Signal Processing and Control}, vol.~71, p.
  103175, 2022. [Online]. Available:
  \url{https://www.sciencedirect.com/science/article/pii/S1746809421007722}
\BIBentrySTDinterwordspacing

\bibitem{Orlandic2021}
L.~Orlandic, T.~Teijeiro, and D.~Atienza, ``{The COUGHVID crowdsourcing
  dataset, a corpus for the study of large-scale cough analysis algorithms.}''
  \emph{Sci Data}, vol.~8, p. 156, Jun 2021.

\bibitem{sharma2020}
N.~Sharma, ``{Coswara -- A Database of Breathing, Cough, and Voice Sounds for
  COVID-19 Diagnosis},'' \emph{arxiv}, 2020.

\bibitem{pizzo2021}
D.~T. Pizzo, ``{IATos: AI-powered pre-screening tool for COVID-19 from cough
  audio samples},'' \emph{arxiv}, 2021.

\bibitem{bagad2020cough}
P.~Bagad, A.~Dalmia, J.~Doshi, A.~Nagrani, P.~Bhamare, A.~Mahale, S.~Rane,
  N.~Agarwal, and R.~Panicker, ``{Cough Against COVID: Evidence of COVID-19
  Signature in Cough Sounds},'' 2020.

\bibitem{2021}
\BIBentryALTinterwordspacing
J.~Andreu-Perez, H.~Perez-Espinosa, E.~Timonet, M.~Kiani, M.~I. Giron-Perez,
  A.~B. Benitez-Trinidad, D.~Jarchi, A.~Rosales, N.~Gkatzoulis, O.~F.
  Reyes-Galaviz, and et~al., ``{A Generic Deep Learning Based Cough Analysis
  System from Clinically Validated Samples for Point-of-Need Covid-19 Test and
  Severity Levels},'' \emph{IEEE Transactions on Services Computing}, p. 1–1,
  2021. [Online]. Available: \url{http://dx.doi.org/10.1109/TSC.2021.3061402}
\BIBentrySTDinterwordspacing

\bibitem{vapnik}
V.~Vapnik, ``{An overview of statistical learning theory},'' \emph{IEEE
  Transactions on Neural Networks}, vol.~10, no.~5, pp. 988--999, 1999.

\bibitem{platt1999}
J.~C. Platt, ``{Probabilistic outputs for support vector machines and
  comparisons to regularized likelihood methods},'' \emph{Advances in Large
  Margin Classifier}, pp. 61--74, 1999.

\bibitem{rabiner}
L.~Rabiner and B.-H. Juang, \emph{{Fundamentals of Speech Recognition}}.\hskip
  1em plus 0.5em minus 0.4em\relax USA: Prentice-Hall, Inc., 1993.

\bibitem{nam}
J.~B. U~Nam, ``{Addressing the Same but different-different but similar problem
  in automatic music classification},'' in \emph{In Proceedings of
  International Symposium in Music Information Retrieval 2001}, 2002.

\bibitem{scheirer}
E.~Scheirer and M.~Slaney, ``{Construction and evaluation of a robust
  multifeature speech/music discriminator},'' in \emph{1997 IEEE International
  Conference on Acoustics, Speech, and Signal Processing}, vol.~2, 1997, pp.
  1331--1334 vol.2.

\bibitem{mulimani}
M.~Mulimani and S.~G. Koolagudi, ``{Acoustic Event Classification Using
  Spectrogram Features},'' in \emph{TENCON 2018 - 2018 IEEE Region 10
  Conference}, 2018, pp. 1460--1464.

\bibitem{tzanetakis}
G.~Tzanetakis and P.~Cook, ``{Musical genre classification of audio signals},''
  \emph{IEEE Transactions on Speech and Audio Processing}, vol.~10, no.~5, pp.
  293--302, 2002.

\bibitem{kumar}
K.~Kumar, C.~Kim, and R.~M. Stern, ``{Delta-spectral cepstral coefficients for
  robust speech recognition},'' in \emph{2011 IEEE International Conference on
  Acoustics, Speech and Signal Processing (ICASSP)}, 2011, pp. 4784--4787.

\bibitem{salim2021}
\BIBentryALTinterwordspacing
H.~Xue and F.~D. Salim, ``{Exploring Self-Supervised Representation Ensembles
  for COVID-19 Cough Classification},'' 2021. [Online]. Available:
  \url{https://arxiv.org/pdf/2105.07566.pdf}
\BIBentrySTDinterwordspacing

\bibitem{toolkit}
\BIBentryALTinterwordspacing
``{s3prl},'' s3prl. [Online]. Available: \url{https://github.com/s3prl/s3prl}
\BIBentrySTDinterwordspacing

\bibitem{tera}
\BIBentryALTinterwordspacing
A.~T. Liu, S.-W. Li, and H.-Y. Lee, ``{TERA: Self-Supervised Learning of
  Transformer Encoder Representation for Speech},'' TERA, 2021. [Online].
  Available: \url{https://arxiv.org/pdf/2007.06028.pdf}
\BIBentrySTDinterwordspacing

\bibitem{librosa}
\BIBentryALTinterwordspacing
``{librosa},'' http://librosa.org/doc/latest/index.html. [Online]. Available:
  \url{http://librosa.org/doc/latest/index.html}
\BIBentrySTDinterwordspacing

\bibitem{loshchilov2017decoupled}
I.~Loshchilov and F.~Hutter, ``{Decoupled weight decay regularization},''
  \emph{arXiv preprint arXiv:1711.05101}, 2017.

\bibitem{rasmussen}
N.~Rasmussen, ``{Cough Sound Analysis for the Evidence of Covid-19. Manuscript
  in preparation},'' manuscript in preparation.

\bibitem{learning}
\BIBentryALTinterwordspacing
``{GitHub - iver56/audiomentations: A Python library for audio data
  augmentation. Inspired by albumentations. Useful for machine learning.}''
  https://github.com/iver56/audiomentations, accessed on Sun, January 02, 2022.
  [Online]. Available: \url{https://github.com/iver56/audiomentations}
\BIBentrySTDinterwordspacing

\bibitem{081rc2}
\BIBentryALTinterwordspacing
\emph{{librosa/librosa: 0.8.1rc2}}, https://doi.org/10.5281/zenodo.4792298,
  accessed on Sun, January 02, 2022. [Online]. Available:
  \url{https://zenodo.org/record/4792298}
\BIBentrySTDinterwordspacing

\bibitem{a2021}
J.~Dinnes, J.~Deeks, S.~Berhane, M.~Taylor, A.~Adriano, C.~Davenport,
  S.~Dittrich, D.~Emperador, Y.~Takwoingi, J.~Cunningham, S.~Beese, J.~Domen,
  J.~Dretzke, L.~Ferrante~di Ruffano, I.~Harris, M.~Price, S.~Taylor-Phillips,
  L.~Hooft, M.~Leeflang, M.~McInnes, R.~Spijker, and A.~Van~den Bruel, ``{How
  accurate are rapid tests for diagnosing COVID-19?}'' \emph{Cochrane Reviews},
  2021.

\bibitem{topol2020}
E.~J. Topol, ``{Is my cough COVID-19?}'' \emph{Lancet Digital Medicine}, 2020.

\bibitem{schulle2021}
H.~Coppock, L.~Jones, I.~Kiskin, and B.~Schulle, ``{COVID-19 detection from
  audio: seven grains of salt},'' \emph{Lancet Digital Health}, 2021.

\end{thebibliography}

\end{document}